\documentstyle[prl,aps,multicol,epsf]{revtex}
\renewcommand{\narrowtext}{\begin{multicols}{2}
\global\columnwidth20.5pc} 
\renewcommand{\widetext}{\end{multicols}
\global\columnwidth42.5pc} \multicolsep = 8pt plus 4pt minus 3pt

\begin{document}

\draft

\title{Electromodulation of the bilayer $\nu$=2 quantum Hall phase
diagram.}

\author{L.Brey$^{1,2}$, E.Demler$^{1}$, and S.Das Sarma$^{1,3}$}

\address{$^1$ Institute for Theoretical Physics, University of California, Santa Barbara, CA 93106}

\address{$^2$ Instituto de Ciencia de Materiales de Madrid (CSIC), 
Cantoblanco, 28049, Madrid, Spain.}
\address{$^3$ Department of Physics, University of Maryland, College Park, MD 20742}


\maketitle

\begin{abstract}
We make a number of precise experimental predictions for observing the various magnetic
phases and the quantum phase transitions between them in the $\nu$=2 bilayer quantum Hall system.
In particular, we analyze the effect of an external bias voltage on the 
quantum phase diagram, finding that a finite bias should readily enable the experimental
observation of the recently predicted novel canted antiferromagnetic phase in transport
and spin polarization measurements.


\end{abstract}

\narrowtext

Recent theoretical work\cite{Zheng,DasSarma1,DasSarma2} predicts the 
existence of a novel canted antiferromagnetic (C) phase in the $\nu$=2
bilayer quantum Hall system under quite general experimental conditions,
and encouraging experimental evidence in its support has recently emerged
through inelastic light scattering spectroscopy\cite{Pellegrini1,Pellegrini2}
and transport measurements\cite{Sawada}. Very recent theoretical works have shown that such a C-phase may exist\cite{Brey1} in a multilayer superlattice system
(with $\nu$=1 per layer), and that in the presence of disorder-induced-interlayer
tunneling fluctuations the C-phase may break up into a rather exotic
spin Bose glass phase\cite{Demler} with the quantum phase transition between
the C-phase and the Bose glass phase being in the same universality class as the
two-dimensional superconductor-insulator  transition in the dirty boson
system.
In this Letter we consider the effect of an external electric 
field induced {\it electromodulation} (through an applied gate bias voltage) of 
the $\nu$=2 bilayer quantum phase diagram.   
Our goal is to provide precise experimental predictions which will
facilitate direct and unambiguous observations of the various magnetic 
phases, and more importantly the quantum phase transitions among them.
We find the effect of a gate bias to be quite dramatic on the $\nu$=2 bilayer quantum phase diagram. 
In particular, a finite gate bias makes the C-phase more
stable which could now exist even in the absence of any interlayer tunneling 
in contrast to the situations considered in references\cite{Zheng,DasSarma1,DasSarma2} 
where the interlayer tunneling induced finite symmetric-antisymmetric gap was crucial
in the stability of the C-phase. Thus, a finite gate bias, according to our
theoretical calculations presented here, has a qualitative effect on the $\nu$=2 bilayer
quantum phase diagram -- it produces a {\it spontaneously interlayer-coherent canted antiferromagnetic
phase} which exists even in the absence of any inter-layer tunneling. The  prediction 
of this spontaneously coherent canted (CC) phase is one of the new theoretical results of this paper.
The theoretical construction of the bilayer $\nu$=2 quantum phase diagram and predicting 
its experimental consequences in the
presence of the bias voltage are   our main results.

The bilayer $\nu$=2 system is characterized by five independent energy scales: the cyclotron energy, $\omega _c$
(we take $\hbar$=1 throughout); the interlayer tunneling energy characterized by $\Delta_{SAS}$, the 
symmetric-antisymmetric energy gap; the Zeeman energy or the spin-splitting $\Delta_z$; the intralayer
Coulomb interaction energy and  the interlayer Coulomb interaction energy. 
The application  of the external electric field adds another independent energy scale, 
the bias voltage, to the problem.
Neglecting the largest ($\omega _c$, which we take to be very large) 
energy scale, one is still left with four independent dimensionless
energy variables to consider in constructing  the $\nu$=2 bilayer quantum phase diagram in the presence of 
finite bias. 
In the absence of any bias 
the quantum phase diagram is surprisingly simple, allowing for
only three qualitatively  different quantum magnetic phases, as established by 
a microscopic Hartree-Fock theory\cite{Zheng,DasSarma1,DasSarma2,Brey1}, a long 
wavelength field theory based on the quantum $O(3)$ nonlinear sigma model\cite{DasSarma1,DasSarma2}, and
a bosonic spin theory\cite{Demler}. These three magnetic phases are the fully spin polarized ferromagnetic phase (F),
which is stabilized  for large values of $\Delta_z$ (or for strong intralayer Coulomb interaction),
the paramagnetic symmetric or spin singlet (S) phase, which is stabilized for large values of $\Delta_{SAS}$
(or for strong interlayer Coulomb interaction), and the intermediate C phase, where 
the electron spins in each layer are tilted away from the external magnetic field direction due to the
competition between ferromagnetic and singlet ordering. 
Note that the S phase is fully pseudospin polarized with $\Delta_{SAS}$, the symmetric-antisymmetric
gap, acting as the effective pseudospin splitting.
The C phase is a true many-body symmetry-broken
phase not existing in the single particle picture 
(and is stabilized by the interlayer antiferromagnetic exchange interaction).
The single particle theory predicts a level crossing and a direct first order transition between the S phase and the F phase 
(nominally at $\Delta_{z} = \Delta_{SAS}$)
as the Zeeman splitting increases. Coulomb interaction creates the new symmetry broken C phase, which prevents any
level crossing (and maintains an energy gap throughout so that there is always a quantized Hall effect)
between F and S phases, and makes all phase transitions in the system continuous second order transitions.
The canted phase is canted in both the spin and the pseudospin space.

One key experimental difficulty in observing the predicted phase transitions (in the absence 
of any external bias voltage) is that a given sample (with a fixed value of $\Delta_{SAS}$, 
determined by the system parameters such as well widths, separations, etc.) is always at a fixed point in the quantum phase 
diagram calculated in references\cite{Zheng,DasSarma1,DasSarma2,Brey1,Demler} because $\Delta_z$, $\Delta_{SAS}$ and
the Coulomb energies are all fixed by the requirement $\nu$=2 and the sample parameters.
Therefore, a given experimental sample in this so-called {\it balanced} condition (i.e. no external bias, equal
electron densities in the two layers on the average) is constrained to lie in the F or C or S phase, and the
only way to see any phase transitions is to make a number of samples with different parameters lying in
different parts of the phase diagram and to investigate and compare properties, as was done in the 
light scattering experiments of references \cite{Pellegrini1,Pellegrini2}. This is obviously an
undesirable situation because what one really wants is to vary an experimental control parameter
(e.g. an external electric field) to tune  the system through the phase boundaries and study the quantum phase
transition instead of studying  different samples. (Theoretically this tuning is easily
achieved by making $\Delta_z$, $\Delta_{SAS}$ and the Coulomb energies continuous variables in the 
phase diagram\cite{Zheng,DasSarma1,DasSarma2,Brey1,Demler}, but experimentally,  of course, this cannot be done.)
In this Letter we show that an externally applied electric field through a gate bias, which takes one
away ({\it off-balance}) from the balanced condition  and introduces\cite{Brey2} unequal layer electron densities is
potentially an extremely powerful experimental tool in studying the $\nu$=2 bilayer quantum phase transitions.
Our results  indicate that using an external gate bias as a 
tuning parameter, a technique already extensively used\cite{Sawada,Suen,Davies} in experimental studies
of bilayer structures, should lead to direct experimental observations of the predicted quantum phases in
$\nu$=2 bilayer systems  and the continuous transitions between them in both transport measurements\cite{Sawada,Suen,Davies}
and in spin polarization measurements through NMR Knight shift experiments\cite{Barret}.

We have used two complementary techniques, the direct Hartree-Fock theory\cite{Zheng,DasSarma1,DasSarma2,Brey1} 
and the effective bosonic spin\cite{Demler}
theory, to evaluate the bilayer $\nu$=2 quantum phase diagram including the effect of a finite bias voltage.
The resulting bias dependent phase diagrams (in the $\Delta_z -\Delta_{SAS}$ space) for the
Hartree-Fock theory and the bosonic spin theory are shown in Figs. 1 and 2, respectively.
Although there are some quantitative differences between the phase diagrams in the two models (to be discussed below),
the main qualitative features are the same: increasing bias voltage enhances the {\it phase space} of the C phase mostly
at the cost of the F phase, and for large enough bias the C phase becomes stable even for 
$\Delta_{SAS}$=0, this CC-phase is spontaneously coherent. We note that the CC-phase
(i.e. the bias induced C-phase along the $\Delta_{SAS}$=0 line) and the C-phase are continuously connected and there 
is no quantum phase transition between them. Note that the S-phase, which is the singlet or the symmetric phase, is
also stabilized for $\Delta_{SAS}$=0 by finite bias effects. This phase (i.e. the S-phase along the $\Delta_{SAS}$=0)
is the spontaneous interlayer coherent symmetric or singlet phase (the CS-phase) and is analogous 
to the corresponding $\nu$=1 spontaneous interlayer  coherent phase studied extensively\cite{Yang}
in the context of the $\nu$=1 bilayer quantum phase diagram. There is, however, a fundamental difference
between the coherent CS phase for our $\nu$=2 bilayer system and the
corresponding\cite{Yang} $\nu$=1 spontaneous interlayer coherent phase; our $\nu$=2 bilayer CS phase can only exist 
under a {\it finite
external bias}  (the same as our CC phase). 
Unlike the corresponding
$\nu$=1 bilayer system\cite{Yang} or the recently studied zero magnetic field bilayer system\cite{Zheng2} there
is no spontaneous breaking of the pseudospin $U(1)$ symmetry
(generated by the interlayer electron density difference)
in our $\nu$=2 coherent bilayer phases which 
can only exist in the presence of an external voltage  bias. We emphasize  that there is no analogy to our canted
phase (C or CC phase) in the corresponding $\nu$=1 bilayer quantum phase diagram\cite{Yang}.

We note that the main difference (cf. Figs.1 and 2) between the Hartree-Fock\cite{Zheng,DasSarma1,DasSarma2,Brey1} theory
and the bosonic spin\cite{Demler} theory
is that the Hartree-Fock theory underestimates the stability of the S phase (compared with the bosonic spin
theory) at small values of $\Delta_z$. This is a real effect and arises from the neglect of
quantum fluctuations in the Hartree-Fock theory which treats the interlayer tunneling as a first order perturbation
correction in the S-phase. The bosonic spin theory is essentially exact for the S-phase and is therefore more
reliable near the C-S phase boundary, particularly for small values of $\Delta_z$ where tunneling effects are
important.

In Fig.3 we show our calculated quantum phase diagrams in the gate voltage ($V_+$) -tunneling ($\Delta_{SAS}$) space
for fixed values of the Zeeman energy $\Delta_z$ (and the Coulomb energies) using both the Hartree-Fock
and the bosonic spin theory. The phase diagrams in the two theories are qualitatively similar, and the interlayer 
coherent phases (CC and CS phases) are manifestly obvious in Fig.3 because the C and the S phases now clearly extend to the
$\Delta_{SAS}$=0 line (the ordinate) for finite bias voltage. In general, the presence of bias therefore allows
for six different quantum magnetic phases in the $\nu$=2 bilayer system:
the usual F,C, and S phases of references \cite{Zheng,DasSarma1,DasSarma2} as well as the purely N\'eel (N)
phase\cite{Zheng,DasSarma1,DasSarma2}
along the $\Delta_z$=0 line in Fig.1 (the F,C,S,N phases are all allowed in the balanced $V_+$=0 situation),
and two new (bias-induced) coherent phases (CC and CS) along the $\Delta_{SAS}$=0 line in Figs. 1-3. The
most important effect of the external bias, which is an  important new prediction of the current paper, is that it allows for a
continuous tuning of the quantum phase of a $\nu$=2 bilayer system within a single gated sample, as is obvious
from Figs.1-3. The predicted quantum phase transitions can now be studied in light scattering\cite{Pellegrini1,Pellegrini2},
transport\cite{Sawada,Suen}, and NMR\cite{Barret} experiments in single gated samples by tuning the  bias voltage
to sweep
through various phases as shown in Figs.1-3.

The last  issue we address here is what one expects to see experimentally in transport and spin polarization measurements, in
sweeping through the phase diagram of Figs.1-3 under an external gate bias. In Fig.4 we show our calculated results
for the variation in the spin polarization of the system as a function of the bias $V_+$ with all the other 
system parameters being fixed. As expected the spin polarization is complete in the F phase and remains a constant as a function
of $V_+$ until it hits the F-C phase boundary where it starts to drop continuously through the C phase,
essentially dropping to zero at  the
the C-S phase boundary, remaining zero in the S-phase. At zero temperature the two 
phase transitions (i.e. F-C and C-S)
are characterized by cusps
in the spin polarization (Fig.4) 
which perhaps  will not be observable in finite temperature experiments. The main features of the calculated spin
polarization as a function of bias, as shown in Fig.4, should, however be readily observable in NMR Knight shift
measurements\cite{Barret}, including possibly the Knight shift difference in the two layers (Fig.4). 
We have also carried out calculations of the interlayer charge imbalance 
(which is zero in the F phase and then rises continuously throughout the C and the S phases reaching 
full  charge polarization for large $V_+$ in the
S phase) as a function of the bias voltage.
There are two cusps in the calculated imbalance as a function of $V_+$, corresponding to the F$\rightarrow$C and the
C$\rightarrow$S phase transitions, which should be experimentally observable.
The calculated imbalance therefore looks almost exactly complementary to the spin polarization results
shown in Fig.4. 
Finally we have also calculated the charged excitation energies within the simple Hartree-Fock
and bosonic theories (assuming no textural excitations such as skyrmions or merons), which
lead to weak cusps in the activation energies at the phase boundaries. Using
the parameters of the samples in ref.\cite{Sawada}, we  conclude from our numerical calculations\cite{note}
that the phase transition being observed in the $\nu=2$ bilayer transport experiments of ref.\cite{Sawada} 
is the transition from the C-phase to the S-phase as a function of the  
density (and {\bf not} from the F -phase to the C -phase as implied in ref.\cite{Sawada}) Neither phase
in ref.\cite{Sawada} is spontaneous interlayer coherent phase (because $\Delta_{SAS}$ is finite in the experiment) in contrast
to the claims of ref.\cite{Sawada}. Our results indicate, however, that it should be possible to see
all three $\nu$=2 quantum phases (F,C and S) in a single gated sample by varying the bias voltage.
We hope that the detailed results presented in this paper will encourage future bilayers 
$\nu$=2 experiments under external gate bias to explore
the predicted rich phase diagram.

We are grateful to A.H.MacDonald and  Z.F.Ezawa for useful discussion.
This work is supported by the National  Science Foundation (at ITP, UCSB). 
LB also acknowledge financial support from grants PB96-0085 and
from the Fundaci\'on Ram\'on Areces. SDS is supported by the US-ONR.

\newpage

\begin{figure}
\centerline{\epsfxsize=8cm 
\epsfbox{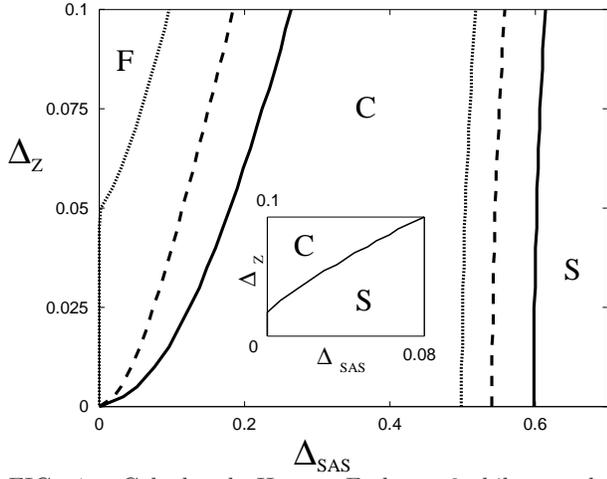}
}
\caption
{Calculated Hartree-Fock $\nu$=2 bilayer phase ($\Delta_{SAS}-\Delta_z$) diagrams
for different bias $V_+$. 
Dot, dashed and continuous lines corresponds to $V_+$=0, 0.5 and 0.65 respectively.
Inset corresponds to the $V_+$=1.42 case.
The length and the energy units are the magnetic length,  $\ell$,
and the intralayer Coulomb energy $e^2/(\epsilon \ell)$. The 
interlayer separation is 1.}
\label{fig1}
\end{figure}
\begin{figure}
\centerline{\epsfxsize=8cm 
\epsfbox{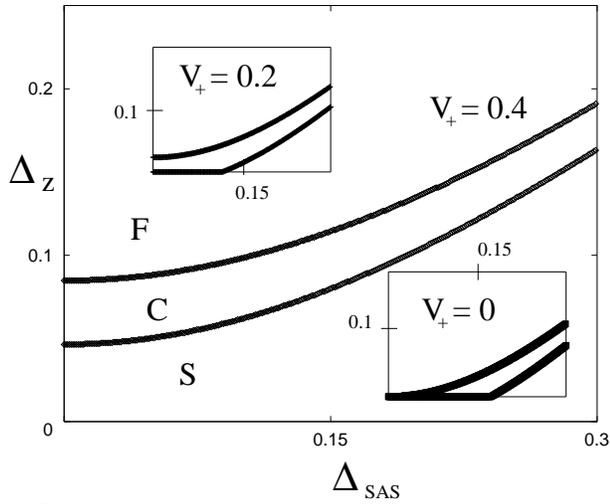}
}
\caption
{ Phase diagram in the bosonic spin theory for different bias voltages. 
All the units and parameters are the same as in Figure 1.
(See ref.[8] for details on the bosonic spin model parameters).
}
\label{fig2}
\end{figure}
\begin{figure}
\centerline{\epsfxsize=8cm 
\epsfbox{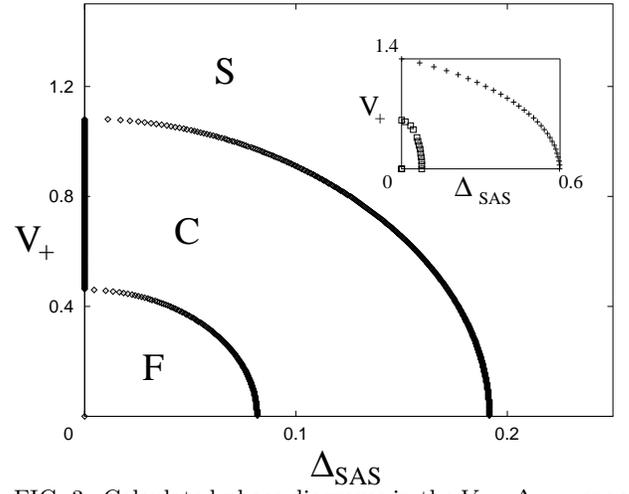}
}
\caption
{Calculated phase diagrams in the $V_+ - \Delta_{SAS}$ space for fixed $\Delta_z$. the main figure:
Bosonic spin phase diagram for $\Delta_z$=0.01 (all other parameters correspond to Fig.2).
Insert: Hartree-Fock phase diagram for $\Delta_z$=0.01 (all other parameters correspond to Fig.1)
}
\label{fig3}
\end{figure}
\begin{figure}
\centerline{\epsfxsize=8cm 
\epsfbox{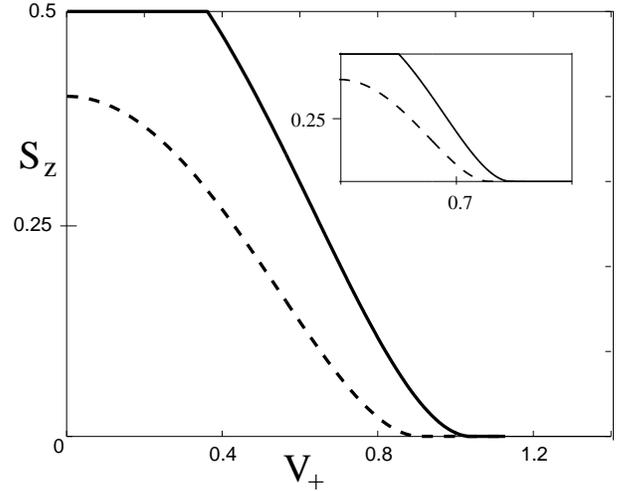}
}
\caption
{Calculated $z$-component of the total spin polarization
 in  each of the layers $\langle S _z \rangle $
as a function  of the bias $V_+$ for fixed
$\Delta_z$ and $\Delta_{SAS}$. 
This quantity is proportional to the Knight shift[12].
Main figure:
solid line Hartree-Fock theory for
$\Delta_{SAS}$=0.05 and dashed line for
$\Delta_{SAS}$=0.1 ( $\Delta_z$=0.01 throughout); all other parameters
correspond to Fig.1. 
Insert: Bosonic spin theory for the same parameters.
}
\label{fig4}
\end{figure}

\widetext

\begin{references}
\bibitem{note}Our numerical Hartree-Fock calculations (not shown here for the sake of brevity) for
the samples of ref.\cite{Sawada} take into account the actual well widths and other
sample parameters given in ref.\cite{Sawada}, and also include realistic Landau level mixing effects.
We find that the $\nu$=2 sample (Fig.4 of ref.\cite{Sawada}) at the balanced density point to be
in the S phase for density $\lesssim 0.5 \times 10 $cm$ ^{-2}$ and in the F phase for 
density $\gtrsim 2.8 \times 10 $cm$ ^{-2}$, and to be in the C-phase in between.
\end{references}
\end{document}